      [1999/12/01 v1.4c Il Nuovo Cimento]
\documentclass{cimento}

\usepackage{graphicx}  

\title{Short GRBs: Rates \& luminosity function implications}

\author{D.~Guetta\from{ins:api}\ETC\thanks{guetta@mporzio.astro.it}}

\instlist{
\inst{ins:api}INAF --- Osservatorio Astronomico di Roma, 
Monteporzio Catone (Roma), Italy}
  
\PACSes{
\PACSit{95.55.Ka}{cosmology-observations}
\PACSit{98.70.Rz}{$\gamma$-ray sources; $\gamma$-ray bursts}
}

\begin{document}

\maketitle

\begin{abstract}
We compare the luminosity function and rate inferred from the
BATSE short hard bursts (SHBs) peak flux distribution with the
redshift and luminosity distributions of SHBs observed by {\it
Swift}/HETE II. The {\it Swift}/HETE II SHB sample is incompatible
with SHB population that follows the star formation rate. However,
it is compatible with a distribution of delay times after the SFR.
This would be the case if SHBs are associated with the mergers of double
neutron star (DNS) systems. DNS may be ``primordial'' or can form 
dynamically by binary exchange interaction in globular clusters 
during core collapse.
The implied SHB rates that we find range from $\sim
8$ to $\sim 30h_{70}^3$Gpc$^{-3}$yr$^{-1}$. This rate is a much
higher than what was previously estimated and, when beaming is
taken into account,  it is comparable to the rate of neutron star
mergers estimated from statistics of binary pulsars. If GRBs are
produced in mergers the implied rate practically guarantees
detection by LIGO II and possibly even by LIGO I.

\end{abstract}

%------------------------------------------------------------------------

\section{Introduction}
In 1993 Koveliotou \cite{ref:kouv93} noticed that the GRB distribution
can be divided into two subsets of long and short bursts with a
dividing duration of 2sec. As short bursts are also harder than
long ones \cite{ref:dez96,ref:kouv96}, they are denoted as Short Hard
Bursts (SHBs). 

The luminosity function (LF) and rate are fundamental quantities
to understand the nature of these objects. Guetta \& Piran (2005) (GP05)
\cite{ref:gp05} have shown how these quantities can be constrained by 
the observed peak flux distribution which is a convolution of these
two unknown functions. 

Afterglows of SHB have only recently been observed by Swift
\cite{ref:geh05,ref:fox05,ref:berg05} and this has lead to the first 
identification of SHB host galaxies and redshift determination. 
In table 1 we report the redshifts and luminosities of these bursts.
As we can see from this table, the average redshift of the short 
bursts is $z\sim 0.4$  much
smaller than the average z of the long GRBs detected by Swift
$z\sim 2.3$.

As BATSE is less sensitive to short bursts than to long ones
\cite{ref:mao94}, even an {\it intrinsic} SHB distribution that follows 
the star formation rate  
(SFR) gives rise to an {\it observed} distribution that is nearer
to us. However the different sensitivity is not enough
to explain the different $z$-distribution ($z$-DF)
between long and short GRBs \cite{ref:gp06}. 
Therefore it is possible that
SHBs do not follow SFR as the long bursts 
and that they are linked to the merger of  double neutron star systems 
(DNS) \cite{ref:eichler89}. In this case
the  SHB rate is given
by the convolution of the star formation rate with the
distribution $P_m(\tau)$ of the merging time delays $\tau$ of the
binary system. These delays reflect the time it takes to the
system to merge due to emission of gravitational radiation.

A delayed SFR distribution (that is intrinsically nearer) gives rise
to an observed distribution that is nearer to us and that can explain
the data for some choices of $P(\tau)$. 
We consider two formation mechanisms. If two massive
stars are born as a binary system, and the system remains bound after
the super nova (SN) explosion of both components, a DNS is
formed (``primordial'' DNS).
Another possibility is that at the moment of star formation the
neutron stars (NSs) are {\it not} in the same binary system, but one
of the NSs is in a binary with a low mass main sequence (MS) star. In
globular clusters (GCs), such binaries are likely to have an exchange
interaction with a single neutron star 
(see \cite{ref:GPZM06} and references within). A significant fraction
($\sim30\%$) of all NS mergers in the Universe may stem from such
dynamically formed systems \cite{ref:GPZM06}.

We show that the predicted $z$-DF is different for the 
two models, so that 
future z-observations may determine which formation channel (if any)
is dominant.

\begin{table}[h]
\caption{The {\it Swift}/HETE II current sample
of SHBs with a known redshift.}
  \centering
\begin{tabular}{|c|c|c|c|c|c|}
\hline
  GRB & 050509b & 050709 & 050724 & 050813 & 051221\\
\hline
  z & 0.22 & 0.16 & 0.257 & 0.7 or 1.80 & 0.5465 \\
\hline
  $L_{\gamma,iso}/10^{51}$erg/sec & 0.14 & 1.1 & 0.17 & 1.9& 3\\
\hline
\end{tabular}
%  \caption{}
\label{table1}
\end{table}

\section{The luminosity function of the BATSE SHB sample}
Our data set and methodology follow \cite{ref:gpw05,ref:gp05,ref:gp06}.
We consider all the SHBs detected while the BATSE onboard trigger
\cite{ref:pa99} was set for 5.5$\sigma$ over background in
at least two detectors in the energy range 50-300keV. These
constitute a group of 194 bursts. We assume the functional form of
the  rate of bursts (but not the amplitude). We then search for a
best fit LF. Using this LF we
calculate the expected distribution of {\it observed} redshifts
and we compare it with the present data. We consider the following
cosmological rates:
\begin{itemize}
    \item (i) A
rate that follows the SFR (We do not expect that this reflects the
rate of SHBs but we include this case for comparison.).
    \item (ii)  A
rate that follows the ``primordial'' DNs merger rate. 
This rate depends on the
formation rate of NS binaries, that one can safely assume follows
the SFR, and on the distribution of merging time delays,
$P_m(\tau)$. This, in turn, depends on the distribution of initial
orbital separation $a$ between the two stars ($\tau\propto a^4$)
and on the distribution of initial eccentricities. Both are
unknown. From the coalescence time distribution of six  double
neutron star binaries \cite{ref:champ04} it seems that
$P_m(\log(\tau))d\log(\tau)\sim$ const, implying $P_m(\tau)\propto
1/\tau$,\cite{ref:piran92}. Therefore, our best guess scenario is a
SBH rate that follows the SFR with a logarithmic time delay
distribution.
    \item (iii)
A rate that follows the SFR with a  delay distribution
$P_m(\tau)d\tau\sim$ const.
    \item (iv) A constant rate (which is
independent of redshift).
\end{itemize}

For the SFR  we employ  $R_{SF2}$ of Porciani \& Madau
\cite{ref:pm01}: In models (ii) and (iii) the rate of SHBs is given
by:
\begin{equation}
\label{rate} R_{\rm SHB}(z) \propto \int_{0}^{t(z)} d\tau R_{\rm
SF2}(t-\tau) P_m(\tau) .
\end{equation}

\begin{table}[h]
\caption{{\bf Table 2:} Best fit parameters Rate(z=0) , $L^*$,
$\alpha$ and $\beta$ and their $1\sigma$ confidence levels for
models (i)-(iv). Also shown are the KS probability ($p_{ks}$) that
the five bursts with a known redshift arise from this
distribution. We show two results for KS tests one with GRB
0508132 at $z= 0.7$ and the other at $z=1.8$. Case ii$_\sigma$
corresponds to case ii with an $L^*$ value lower by $1 \sigma$
than the best fit one. Other parameters have been best fitted for
this fixed number.}
%------------------
\begin{tabular}{|c|c|c|c|c|c|c|}
 \hline
& Rate(z=0) & $L^*$ & $\alpha$ & $\beta$& $p_{KS}$ & $p_{KS}$ \\
   & $Gpc^{-3} yr^{-1}$ & $10^{51}$ erg/sec &   & &(z=0.7)&(z=1.8) \\
  \hline
i & $0.11^{+0.07}_{-0.04}$ & $4.6^{+2.2}_{-2.2}$
&$0.5^{+0.4}_{-0.4} $ &
$1.5^{+0.7}_{-0.5} $&$<0.01$&$<0.01$ \\
 ii &  $0.6^{+8.4}_{-0.3}$ & $2^{+2}_{-1.9}$ &
$0.6 ^{+0.4}_{-0.4}$ & $2 \pm 1$& 0.05 &0.06\\
ii$_\sigma$ & $10^{+8}_{-5}$ & 0.1 & $0.6^{+0.2}_{-0.4}$ &
$1\pm 0.5 $& 0.22 &0.25  \\
 iii & $30^{+50}_{-20}$ & $0.2^{+0.5}_{-0.195}$  &
$0.6^{+0.3}_{-0.5} $  & $1.5^{+2}_{-0.5} $&0.91 & 0.91\\
iv & $8^{+40}_{-4}$ & $0.7^{+0.8}_{-0.6}$  &
$0.6^{+0.4}_{-0.5} $  & $2^{+1}_{-0.7} $& 0.41 & 0.41\\
 \hline
\end{tabular}
\end{table}
%\caption{

Following  Schmidt \cite{ref:sch01} (see also \cite{ref:gpw05,ref:gp05}) we
consider a broken power law peak LF  with lower
and  upper limits, $1/\Delta_1$  and $\Delta_2$, with power law
indeces, $\alpha$, $\beta$ and luminosity break $L^*$.

We use $\Delta_{1,2}=(30,100)$ \cite{ref:gp05}. In \cite{ref:gp06}
(denoted GP06 hereafter) we show that both limits are chosen in
such a way that a very small fraction (less than 1\%) of the {\it
observed} bursts are outside these range. Hence one cannot infer
anything from the observations on the LF in this
range. Comparing the predicted distribution with the one observed
by BATSE we obtain, the best fit parameters of each model and
their standard deviation. The results are shown in table 2 and in
figs 1 and 2 of GP06.

\section{A Comparison with the current {\it Swift}-HETE II SHB sample}

We can  derive now the expected $z$-DF of the
observed bursts' population in the different models.

We assume that the minimal peak flux for detection for {\it Swift}
is $ \sim 1$ ph/cm$^{2}$/sec like BATSE (note the different
spectral windows of both detectors which makes {\it Swift}
relatively less sensitive to short  bursts).

\begin{figure}[b]
  \includegraphics[height=.3\textheight]{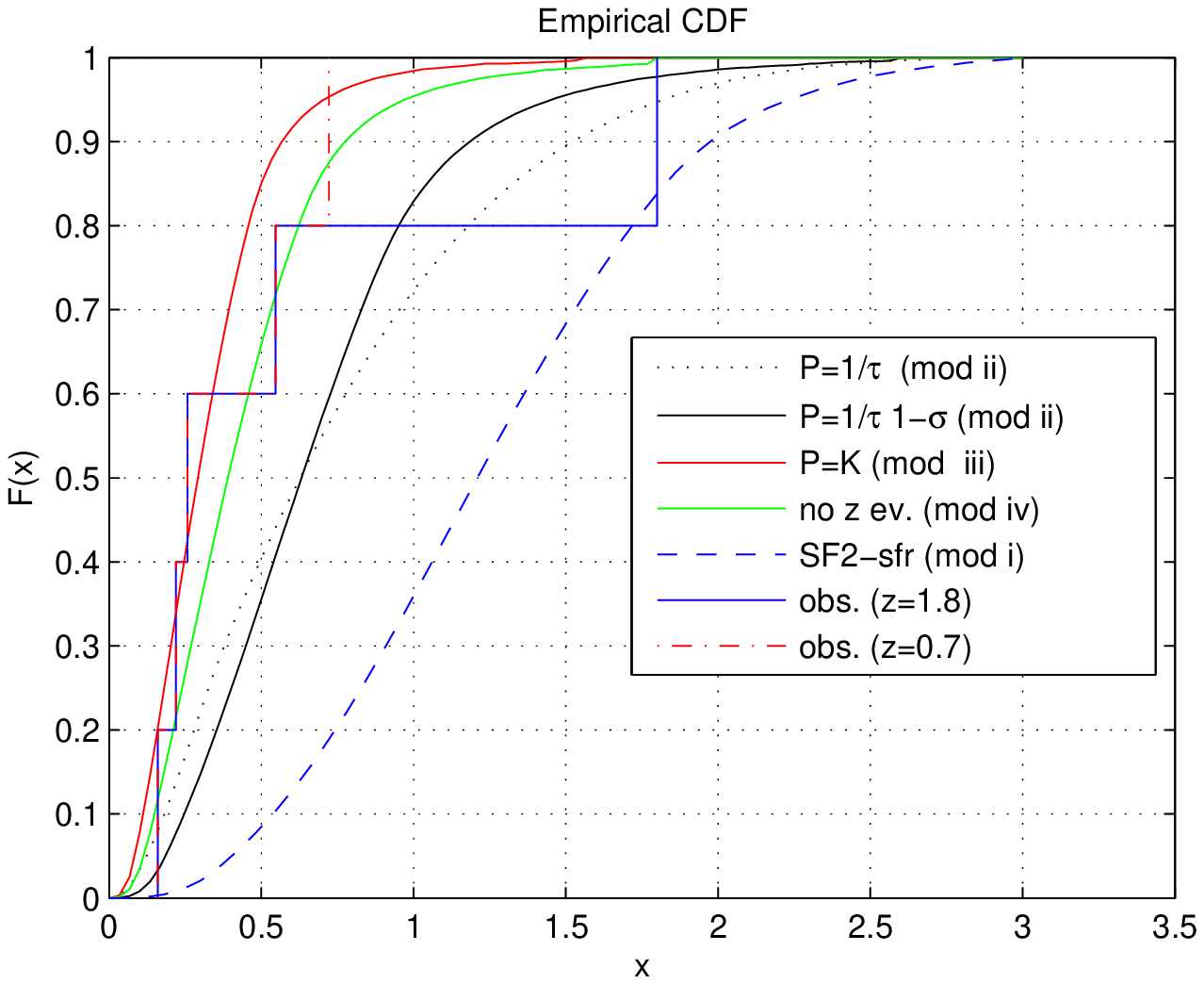}
 \includegraphics[height=.3\textheight]{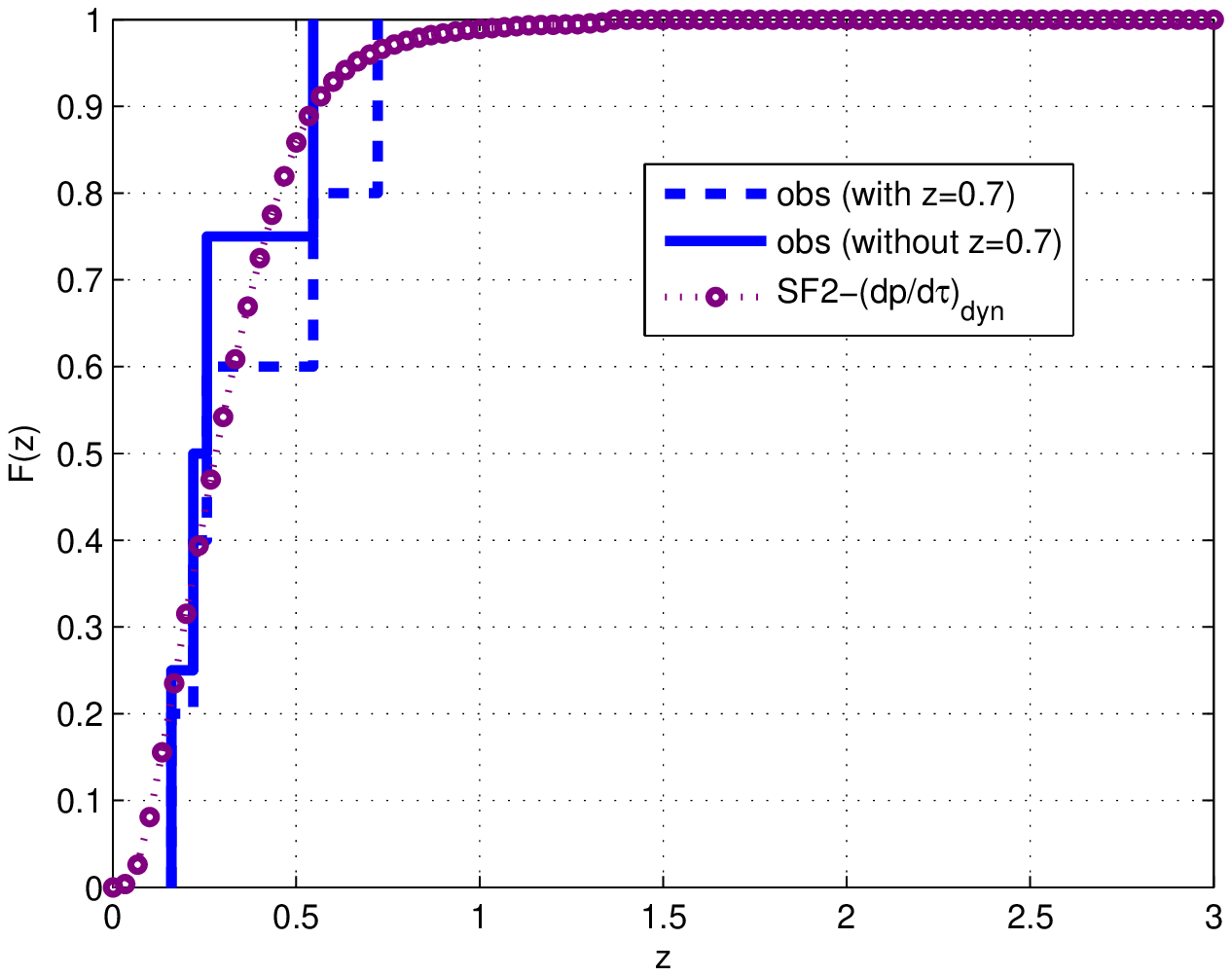}
  \caption{\label{fig1}Left panel:   A comparison between
the expected integrated {\it observed} $z$-DF of
SHBs for
 models (i)-(iv) and (ii$_\sigma$) and the distribution of
known redshifts of SHBs. Right panel:  A comparison between
the expected integrated {\it observed} $z$-DF of
SHBs for dynamically formed DNS }
\end{figure}

Fig.\ref{fig1} (left panel)  depicts the expected {\it observed} integrated
$z$-DF of SHBs in the different models. As
expected, a distribution that follows the SFR, (i), is ruled out
by a KS test with the current five bursts ($p_{KS} <1$\%). This is
not surprising as other indications, such as the association of
some SHBs with elliptical galaxies suggest that SHBs are not
associated with young stellar populations.

A distribution that follows the SFR with a constant logarithmic
delay distribution, (ii), is marginally consistent with the data
($p_{KS} \sim 5-6$\%). The observed bursts are nearer (lower
redshift) than expected from this distribution. If we use the
Rowan-Robinson SFR \cite{ref:gp05}, rather than SF2 of Porciani-Madau,
the situation is even more promising $p_{KS} \sim 10$\% (for
either z=0.7 or z-1.8). When we move to a distribution that is
1$\sigma$ away from the best fit distribution we find $p_{KS}\sim
22-25$\% and even higher for the RR SFR. Thus the suggestion
\cite{ref:GalYam05a} that the observed sample rules out the NS merger
model (with a logarithmic merger time distribution) was somewhat
premature.  Note however, that the local rate with this model
(ii$_\sigma$) is sixteen times larger than the rate of the best
fit model, (ii). This reflects the large flexibility in modeling
the peak flux distribution.

To demonstrate the flexibility of the data we have considered two
other time delay distributions. A uniform time delay distribution
(iii) and an overall constant SHB rate (iv). Both models are
compatible with the BATSE SHB distribution and with the sample of
SHBs with a known redshift ($p_{KS}\sim $ 80\% and 40\%
respectively.). This result is not surprising. The BATSE peak flux
distribution depends on two unknown functions, the rate and the
LF. There is enough freedom to chose one function
(the rate) and fit for the other.

In all  models compatible with the five bursts with a known
redshift, the {\it intrinsic} SHB rate is pushed towards lower
redshifts. The  inferred present rates, $\sim 30$, $\sim 8$ and
$\sim 10h_{70}^3$Gpc$^{-3}$yr$^{-1}$ for models (iii), (iv) and
(ii$_\sigma$) respectively, are larger by a factor ten to fifty
than those estimates earlier (assuming that SBHs follow the SFR
with a logarithmic delay with the best fit parameters GP05).  The
corresponding ``typical" luminosities, $L^*$, ranges from 0.1 to
0.7  $\times 10^{51}$erg/sec.

Let us consider now the possibility that short bursts come from dynamically
formed DNS \cite{ref:hopman06}.
In this case, the delay time $\tau$ is a sum of the time $t_{cc}$ until the
dynamical formation of a DNS during core-collapse, and the time
$t_{\rm GW}$ until the DNS merges. The resulting
delay function $(dp/d\tau)_{\rm dyn}$ of dynamically formed DNSs
is given by \cite{ref:hopman06}

\begin{equation}\label{eq:Ftau}
\left({dp\over d\tau}\right)_{\rm dyn}= {d\over d\tau}\int_0^{\tau}dt_{cc}{dp_{cc}\over t_{cc}}(t_{cc})\int_0^{\tau-t{cc}}dt_{\rm GW} {dp_{\rm GW}\over t_{\rm GW}}(t_{\rm GW}).
\end{equation}

The resulting delay function is shown in Fig.2  of \cite{ref:hopman06}.
Fig.1 (right panel) 
shows the comparison between the expected and observed integrates 
$z$-DF. Our model for dynamical mergers fits the observed 
SHBs distribution much better ($p_{ks}\sim 70\%-80\%$)
than the model for primordial distribution 
with $P(\tau)\sim 1/\tau$ ($p_{ks}\sim 5\%-20\%$) .

\section{Conclusions and Implications}

 We have repeated the analysis of fitting the BATSE SHB data to a
model of the luminosities and rates distributions. Our best fit
logarithmic distribution model is similar to the best fit
logarithmic model presented in GP05. A main new ingredient of this
work is the fact that we consider several other models. We confirm
our earlier finding that the BATSE data allows a lot of
flexibility in the combination of the rates and luminosities.

A second new ingredient of this work is the comparison of  the
best fit models to the small sample of five {\it Swift}/HETE II
SHBs. The {\it Swift}/HETE II data gives a new constraint. This
constraint favors a population of SHBs with a lower intrinsic
luminosity and hence a nearer {\it observed} $z$-DF. 
It implies a significantly higher local SHB rate - a
factor of ten to fifty higher than earlier estimates. The new
observations of Swift show that the SHBs are nearer than what was
expected before and therefore, their luminosity is lower and their
local rate is higher. We stress that this new result was within
the 1$\sigma$ error of the model presented in GP05, which had a
very large range of allowed local rates and typical luminosities.

Provided that the basic model is correct and we are not mislead by
statistical (small numbers), observational (selection effects and
threshold estimates) of intrinsic (two SHB population) factors we
can proceed and compare the inferred SHB rate with the
observationally inferred rate of NS-NS mergers in our galaxy
\cite{ref:na91}. This rate was recently reevaluated with the discovery
of PSR J0373-3039 to be rather large as $80^{+200}_{-66}$/Myr.
Although the estimate contains a fair amount of uncertainty
\cite{ref:ka04}. If we assume that this rate is typical and that the
number density of galaxies is $\sim 10^{-2}$/Mpc$^{3}$, we find a
merger rate of $800^{+2000}_{-660}$/Gpc$^{3}$/yr. Using a beaming
factor of 30-50 for short bursts \cite{ref:berg05} this rate implies a
total merger rate of $\sim 240-1500$/Gpc$^{3}$/yr for the three
cases (iii), (iv) and (ii$_\sigma$). The agreement between the
completely different estimates is surprising and could be
completely coincidental as both estimates are based on very few
events.

If correct these estimates are excellent news  for gravitational
radiation searches, for which neutron star mergers are prime
targets. They imply that the recently updates high merger rate,
that depends mostly on one object, PSR J0737+3039, is valid. These
estimate implies one merger event within $\sim 70$Mpc per year and
one merger accompanied with a SHB within $\sim 230$Mpc. These
ranges are almost within the capability of LIGO I and certainly
within the capability of LIGO II. If correct these estimates of
the rate are excellent news  for gravitational radiation searches,
for which neutron star mergers are prime targets.

We have considered also the possibility that the DN system is formed 
dynamically and we have shown  that the $z$-DF of 
short GRBs expected from {\it dynamical} formation and subsequent merger 
of DNSs is markedly different from that expected from primordial DNS 
mergers (under the assumption that the distribution of delay times for 
primordial DNSs follows $(dp/d\tau)_{\rm prim}\propto 1/\tau$). The large
time for core collapse shifts the distribution of dynamically formed
DNS mergers to low redshift. The observed $z$-DF of 
short GRBs is consistent with that expected for mergers of dynamically formed
DNSs, and appears to be inconsistent with that expected for primordial
DNS mergers if the best fit parameters are considered
We note, however, that current data do not
allow to rule out a $z$-DF consistent with that expected for
primordial DNS mergers, since redshifts were obtained only for a minority
of the detected short GRBs. This may be due to a bias against obtaining 
redshift information for high redshift (faint) short GRBs (see Fig.4,
table~1 and discussion at the end of \S3.3 of \cite{ref:hopman06}). 
Future observations should 
allow to better constrain the $z$-DF of short GRBs, and thus
to differentiate between models. For example, detection of only a few high 
redshift ($z>2$) short GRBs would severely constrain the contribution of
dynamically formed DNSs.

To conclude we stress that we have assumed that the luminosity
function has a lower limit of $L^*/30$. This was just because even
if such a limit does not exists weaker bursts would be barely
detected. The current peak flux distribution of BATSE burst cannot
confirm (or rule out) the existence of such population (note
however, that Tanvir et al., \cite{ref:tavnir05} suggest that such a
population exists on the basis of the angular distribution of
BATSE SHBs). If such weak bursts exist then, of course, the
overall merger rate will be much larger \cite{ref:n05}. Such events
will provide such a high rate that soon LIGO I will begin to
constrain this possibility.


\begin{thebibliography}{0}


\bibitem{ref:kouv93}
\BY{Kouveliotou~C. et al.}, 
\IN{Astrophysical Journal}{413}{1993}{L101}.

\bibitem{ref:dez96}
\BY{Dezalay~J. P. et al.},
\IN{Astrophysical Journal}{471}{1996}{L27}.

\bibitem{ref:kouv96}
\BY{Kouveliotou~C. et al.}, 1996, in {\it Gamma-ray bursts,
Proceedings of the 3rd Huntsville Symposium. AIP Conference
proceedings series, 1996, vol. 384, edited by Kouveliotou et al.,
ISBN: 156966859, p.42.}

\bibitem{ref:gp05} \BY{Guetta~D. \& Piran~T.}
\IN{Astronomy \& Astrophysics}{435}{2005}{421}

\bibitem{ref:geh05}\BY{Gehrels~N. et al.}
\IN{Nature}{437}{2005}{851}

\bibitem{ref:fox05}\BY{Fox~D. B. et al.}
\IN{Nature}{437}{2005}{845}

\bibitem{ref:berg05} \BY{Berger~E. et al.}
\IN{Nature}{438}{2005}{988}


\bibitem{ref:mao94}
\BY{Mao~S., Narayan~R. \& Piran~T.}
\IN{Astrophysical Journal}{420}{1994}{171}.


\bibitem{ref:gp06}
\BY{Guetta~D. \& Piran~T.}
\IN{Astronomy \& Astrophysics}{453}{2006}{823}
 


\bibitem{ref:eichler89}
 \BY{Eichler~D., Livio~M., Piran~T. \& Schramm~D.}
\IN{Nature}{340}{1989}{126}

\bibitem{ref:GPZM06} \BY{Grindlay~J., Portegies Zwart~S. F. \& 
McMillan S.}
\IN{Nature Physics}{2}{2006}{116}

\bibitem{ref:gpw05}
\BY{Guetta~D., Piran T. \& Waxman~E.}
\IN{Astrophysical Journal}{619}{2005}{412}.

\bibitem{ref:pa99}
\BY{Paciesas~W.S. et al.}
\IN{Astrophysical Journal Supplement}{122}{1999}{465}. 

\bibitem{ref:champ04}
\BY{Champion~D.J. et al.}
\IN{Mountly Notices RAS}{350}{2004}{L61}.

\bibitem{ref:piran92} 
\BY{Piran~T.}
\IN{Astrophysical Journal}{389}{1992}{L83}.

\bibitem{ref:pm01}
\BY{Porciani~C. \& Madau~P.}
\IN{Astrophysical Journal}{548}{2001}{522}.

\bibitem{ref:sch01}
\BY{Schmidt~M.}
\IN{Astrophysical Journal}{559}{2001}{L79}.

\bibitem{ref:GalYam05a} 
\BY{Gal-Yam~A. et al.}
astro-ph/0509891.

\bibitem{ref:hopman06}
\BY{Hopman~C., Guetta~D., Waxman~E. \& Portegies Zwart~S.}
\IN{Astrophysical Journal}{643}{2006}{L91}

\bibitem{ref:na91}
\BY{Narayan~R., Piran~T. \& Shemi~A.}
\IN{Astrophysical Journal}{379}{1991}{L17}.

\bibitem{ref:ka04}
\BY{Kalogera~V. et al.}
\IN{Astrophysical Journal}{614}{2004}{L137} 

\bibitem{ref:tavnir05} 
\BY{Tanvir~N. et al.}
\IN{Nature}{438}{2005}{991}

\bibitem{ref:n05} 
\BY{Nakar~E., Gal Yam.~A. \& Fox~D.}
astro-ph/0511254

\end{thebibliography}
\end{document}